\documentclass[a4paper,12pt]{article}
\usepackage{graphicx}
\newcommand{\be}{\begin{equation}}
\newcommand{\ee}{\end{equation}}
\newcommand{\bea}{\begin{eqnarray}}
\newcommand{\eea}{\end{eqnarray}}
\newcommand{\beqn}{\begin{eqnarray}}
\newcommand{\eeqn}{\end{eqnarray}}
\newcommand{\ba}{\begin{array}}
\newcommand{\ea}{\end{array}}

\newcommand{\noi}{\noindent}
\newcommand{\grcl}{{\tt GRACE-loop$\;$}}
\newcommand{\ra}{\rightarrow}
\newcommand{\epem}{e^+ e^-}
\newcommand{\epemt}{$e^+ e^- \;$}

\newcommand{\eettht}{$e^+ e^-\ra t \bar{t} H\;$}

\newcommand{\eennht}{$\epem \ra \nu \bar{\nu} H \;$}
\newcommand{\eennh}{$\epem \ra \nu \bar{\nu} H$}

\newcommand{\eezht}{$\epem \ra Z H \;$}
\newcommand{\eezhh}{$\epem \ra Z  H H$}
\newcommand{\eezhht}{$\epem \ra Z H H\;$}

\def\sm{${\cal{S}} {\cal{M}}\;$}
\def\sms{${\cal{S}} {\cal{M}}$}

\newcommand{\ordalf}{{\cal O}(\alpha)}
\newcommand{\ordalft}{${\cal O}(\alpha)\;$}

\newcommand{\beq}{\begin{equation}}
\newcommand{\eeq}{\end{equation}}

\newcommand{\np}{Nucl.\,Phys.\,}
\newcommand{\pl}{Phys.\,Lett.\,}
\newcommand{\pr}{Phys.\,Rev.\,}

\begin{document}

\begin{titlepage}

\vspace*{0.1cm} \rightline{LAPTH-994}
\vspace*{0.1cm}\rightline{KEK-CP-142}


\vspace{1mm}
\begin{center}

{\Large{\bf Full ${\cal O}(\alpha)$ electroweak corrections to
double Higgs-strahlung  at the linear collider}}

\vspace{.5cm}

G. B\'elanger${}^{1)}$, F. Boudjema${}^{1)}$, J.
Fujimoto${}^{2)}$, T. Ishikawa${}^{2)}$, \\ T. Kaneko${}^{2)}$,
Y.~Kurihara${}^{2)}$, K. Kato${}^{3)}$, Y. Shimizu${}^{2)}$
\\

\vspace{4mm}

{\it 1) LAPTH${\;}^\dagger$, B.P.110, Annecy-le-Vieux F-74941,
France.}
\\ {\it
2) KEK, Oho 1-1, Tsukuba, Ibaraki 305--0801, Japan.} \\
{\it 3)
Kogakuin University, Nishi-Shinjuku 1-24, Shinjuku, Tokyo
163--8677, Japan.} \\

\vspace{10mm} \abstract{ We present the full ${{\cal O}}(\alpha)$
electroweak radiative corrections to  the double Higgs-strahlung
process \eezhht.  The computation is performed with the help of
{\tt GRACE-loop}. After subtraction of the initial state QED
radiative corrections, we find that the genuine weak corrections
in the $\alpha$-scheme  are small for Higgs masses and energies
where this cross section is largest and is most likely to be
studied. These corrections decrease with increasing energies
attaining about $\sim -10\%$ at $\sqrt{s}=1.5$TeV. The full
${{\cal O}}(\alpha)$ correction on the other hand is quite large
at threshold but small at energies around the peak. We also study changes
in the shape of the invariant mass of the Higgs pair which has
been shown to be a good discriminating variable for the
measurement of the triple Higgs vertex in this reaction.}

\end{center}

\vspace*{\fill} $^\dagger${\small URA 14-36 du CNRS, associ\'ee
\`a l'Universit\'e de Savoie.} \normalsize
\end{titlepage}

\section{Introduction}
One of the primary goals of the first phase of a future $e^+e^-$
linear collider\cite{NLC-report,tesla-report,GLC-report}, LC, is a
detailed precision study of the properties of the Higgs. According
to the latest indirect precision data\cite{mhlimit-03-2003}, one
expects a rather light Higgs with a mass below the $W$ pair
threshold. This is also within  the range predicted for the
lightest Higgs of the minimal supersymmetric model(MSSM). The LHC
will be able to furnish a few measurements on the couplings of the
Higgs to fermions and gauge bosons\cite{higgsproperties-lc-lhc}
but the most precise measurements will be performed in the clean
environment of  the linear collider. In order for these precision
measurements to be feasible at the LC, the various observables
need to be well under control theoretically. Nevertheless,
although the branching ratios of the Higgs have been computed with
great precision it is only during the last year that the one-loop
electroweak radiative corrections to the main production channel
at the LC, \eennh, has been achieved\cite{eennhradcor2002,
eennhletter,Dennereennh1}. Even more recent
is the calculation\cite{eetthgrace,eetthdenner,eetthchinese} of
$e^+e^- \rightarrow t\bar{t}H$ which is crucial for the extraction
of the important $t \bar t H$ vertex. Another important issue
concerns the reconstruction of the Higgs potential which is the
cornerstone of the electroweak symmetry breaking. Although
measurements of the Higgs mass and its couplings to the fermions
and bosons may give some information on this
potential\cite{me-3h-2hdm}, a more direct probe is through the
measurement of  the Higgs self-couplings. In particular the
tri-linear Higgs self-coupling, $HHH$, may be accessed in double
Higgs production. Especially for the light Higgs mass this is a
daunting task at the LHC\cite{all-lhc-3h,Baur3h-lhc-lc}. In \epemt the most
promising channel in the first stage of a LC, double
Higgs-strahlung, \eezhht \cite{trilinearHiggsSM,ChopinHiggs},
seems to be the most promising. The aim of this letter is to give
the  full electroweak correction to this process, thus adding to
the precision calculations of the Higgs profile at the LC.

The cross sections for \eezhht are rather small reaching about
$0.2$fb at centre of mass energies around $500$GeV and for Higgs
masses in the interesting range $110<M_H<130$GeV as suggested by
supersymmetry. However detailed
simulations\cite{zhhexpfrancais} including the issue of
backgrounds have shown that a precision of about $10\%$ on the
total cross section can be achieved with a collected luminosity of
$1$ab$^{-1}$ especially if a neural networks analysis is
performed. Other simulations\cite{hhhbattagliaboos,jlc-3h} have
shown that by using some discriminating kinematics variables,
namely the invariant mass of the $HH$ system, one can  further
improve the extraction of the Higgs self-coupling $\lambda_{HHH}$.
It is important to point out that although the foreseen
experimental precision on the cross section is about $10\%$, the
knowledge of the complete \ordalft corrections is mandatory for
this process. Indeed we already know  that the radiative
corrections to the Higgs self-couplings alone, which is just a
small subset of the full electroweak corrections, receive a
$M_t^4$ correction and are therefore  potentially large, with $M_t$
being the top mass. The leading correction to the tree-level
$\lambda_{HHH}^{(0)}$ coupling writes\footnote{This can, for example, be
extracted from the analysis in
\cite{Holliksusyselfcouplings}. We have also made an independent
derivation.}
\beqn
\label{mt4hhh}
\lambda_{HHH}=\lambda_{HHH}^0\left( 1-\frac{\alpha}{\pi s_W^2}
\frac{M_t^4}{M_W^2 M_H^2}\right), \quad s_W^2=1-M_W^2/M_Z^2.
\eeqn
\noi $M_W$ and $M_Z$ are, respectively, the $W$ mass and $Z$ mass
and $\alpha$ the fine structure constant. For $M_H=120$GeV, this
constitutes as much as $-10\%$ correction to the tri-linear Higgs
coupling. Although the contribution from diagrams from  the triple
Higgs vertex to the full process are not the dominant ones, this
correction alone can amount to about $-5\%$ to the total cross
section\cite{trilinearHiggsSM}.

\section{Grace-Loop and the calculation of \eezhht}
\subsection{Checks on the one-loop result}
 Our computation is performed with the help of {\tt GRACE-loop}\cite{nlgfatpaper}.
This is a code for the automatic generation and calculation of the
full one-loop electroweak  radiative corrections in the \sms. The
code  has  successfully reproduced the results of a host of
one-loop $2\ra 2$ electroweak processes\cite{nlgfatpaper}. {\tt
GRACE-loop} also provided the first results on the full one-loop
radiative corrections to $e^+ e^- \ra \nu \bar{\nu} H$
\cite{eennhradcor2002,eennhletter} and more recently on \eettht\cite{eetthgrace}.
Both these calculations  have now  been confirmed by an
independent calculation\cite{Dennereennh1,eetthdenner}. For all
electroweak processes we adopt the on-shell renormalisation scheme
according to\cite{eennhletter,nlgfatpaper,kyotorc}. For each
process some stringent consistency checks are performed. The
results are checked by performing three kinds of tests at some
random points in phase space. For these tests to be passed one
works in quadruple precision. Details of how these tests are
performed are given in\cite{eennhletter,nlgfatpaper}. Here we
only describe the main features of these tests. \\
\noi {\it i)}  We first check the ultraviolet finiteness of the
results. This test applies to the whole set  of the virtual
one-loop diagrams. In order to conduct this test we regularise any
infrared divergence by giving the photon a fictitious mass (for
this calculation we set this at $\lambda=10^{-21}$GeV). In the
intermediate step of the symbolic calculation dealing with loop
integrals (in $n$-dimension), we extract the regulator constant
$C_{UV}=1/\varepsilon -\gamma_E+\log 4\pi$, $n=4-2 \varepsilon$
and treat this as a parameter. The ultraviolet finiteness test is
performed by varying the dimensional regularisation parameter
$C_{UV}$. This parameter could then be set to $0$ in further
computation. Quantitatively for the process at hand, the
ultraviolet finiteness test gives a result that is stable over
$29$ digits when one varies the dimensional regularisation
parameter $C_{UV}$.\\

\noi {\it ii)} The test on the infrared finiteness is performed by
including both the loop and the soft bremsstrahlung contributions
and checking that there is no dependence on the fictitious photon
mass $\lambda$. The soft bremsstrahlung part  consists of a soft
photon contribution where the external photon is required to have
an energy $k_{\gamma}^0 < k_c\ll E_b$. $E_b$ is the beam energy.
This part factorises
and can be dealt with analytically. For the QED infrared finiteness
test we also find results that are stable over $23$ digits when
varying the fictitious photon mass $\lambda$.

\noi {\it iii)} A crucial test concerns the gauge parameter
independence of the results. Gauge parameter independence of the
result is performed through a set of five gauge fixing parameters.
For the latter a generalised non-linear gauge fixing
condition\cite{nlgfatpaper} has been chosen,
\beqn
\label{fullnonlineargauge} {{\cal L}}_{GF}&=&-\frac{1}{\xi_W}
|(\partial_\mu\;-\;i e \tilde{\alpha} A_\mu\;-\;ig c_W
\tilde{\beta} Z_\mu) W^{\mu +} + \xi_W \frac{g}{2}(v
+\tilde{\delta} H +i \tilde{\kappa} \chi_3)\chi^{+}|^{2} \nonumber \\
& &\;-\frac{1}{2 \xi_Z} (\partial.Z + \xi_Z \frac{g}{ 2 c_W}
(v+\tilde\varepsilon H) \chi_3)^2 \;-\frac{1}{2 \xi_A} (\partial.A
)^2 \;.
\eeqn
The $\chi$ represents the Goldstone. We take the 't Hooft-Feynman
gauge with $\xi_W=\xi_Z=\xi_A=1$ so that no ``longitudinal" term
in the gauge propagators contributes. Not only this makes the
expressions much simpler and avoids unnecessary large
cancellations, but it also avoids the need for high tensor
structures in the loop integrals. The use of the five parameters,
$\tilde{\alpha}, \tilde{\beta}, \tilde{\delta}, \tilde{\kappa},
\tilde\varepsilon $ is not redundant as often these parameters
check complementary sets of diagrams.  Let us also point out that
when performing this check we keep the full set of diagrams
including couplings of the Goldstone and Higgs to the electron for
example, as will be done for the process under consideration. Only
at the stage of integrating over the phase space do we switch
these negligible contributions off. Here, the gauge parameter
independence checks give results that are stable over $27$ digits
(or better) when varying any of the non-linear gauge fixing
parameters.

\subsection{The five point function}
The gauge invariance check is also a very powerful check as
concerns the reduction of the various tensor integrals down to the
scalar integrals of rank $N$. For $N<5$ we perform this reduction in terms of
Feynman parameters as detailed in our review\cite{nlgfatpaper}. As
for the scalar $N=3,4$ integrals we use the {\tt FF}
package\cite{ff}. The implementation of the scalar five-point
function is done exactly as in our previous paper on
\eennht\cite{eennhletter}. On the other hand we have improved the
algorithm for the reduction of the tensor integrals for the five
point function. A general five point function is constructed out
of  tensor five point functions and  is written as

\beqn
T^{(5)}=\int \frac{d^n l}{(2\pi)^n} \; \frac{N(l)}{D_0 D_1 \cdots
D_{4}}=G^{ \mu \nu \cdots \rho} \; \int \frac{d^n l}{(2\pi)^n} \;
\frac{\overbrace{l_\mu l_\nu \cdots l_\rho}^{M}}{D_0 D_1 \cdots
D_{4}}, \quad M \leq 5,
\eeqn
\noi where

\beqn
\label{didi} D_0=l^2-M_0^2 \quad D_i=(l+s_i)^2-M_i^2, \quad
s_i=\sum_{j=1}^{i} p_j, \qquad i=1,2,3,4.
\eeqn
\noi $M_i$ are the internal masses, $p_i$ the set (of linearly
independent) incoming momenta, $l$ the loop momentum and $G^{\mu
\nu\cdots \rho}$ a tensor that involves the external momenta, the
metric or anti-symmetric
tensors. The scalar integral has  $N(l)=1$.  \\
Because the set $s_i$  forms a basis for vectors in 4-dimensional
space, introducing the Gram matrix $A_{ij} = s_i.s_j$ one has the
identities
\begin{eqnarray}
\label{l2}
   g^{\mu\nu} = \sum_{i,j=1}^{4} s_{i}^\mu A_{ij}^{-1} s_{j}^\nu,
   \quad l^\mu &=& \sum_{i, j} s_i^\mu A^{-1}_{ij} (l . s_j).
\end{eqnarray}
We then rewrite an occurrence of $l^2$ in $N(l)$ in terms of
$D_0$. For a term with $l_\mu$ we use Eq.(\ref{l2}) and re-express
$l.s_j$ as a difference of $D_j$ and $D_0$ (and loop momenta
independent terms). Using this trick one is able to write
\begin{eqnarray*}
    N(l) &=& \sum_{\alpha=0}^4 E_\alpha(l) D_\alpha + F,
\end{eqnarray*}
where \(F\) is independent of the loop momentum and corresponds
then to the scalar five-point function which we treat as
in\cite{eennhletter}. The terms proportional to $D_\alpha$
correspond to a tensor four-point function of rank-$M-1$ or $M-2$,
starting from a rank-$M$ in the original $T^{(5)}$. We use {\tt
REDUCE} for extracting \(E(l)\) and \(F\) analytically. This new
algorithm can provide much shorter matrix elements compared with
the previous one, which used the identity
\begin{eqnarray*}
l^2 &=& \sum_{i,j} (l.s_i) A^{-1}_{i,j} (l.s_j),
\end{eqnarray*}
instead of $l_\mu$ in Eq.(\ref{l2}).

\subsection{Input parameters}
Our input parameters for the calculation of $e^+e^-\rightarrow Z H
H$ are the following. We will start by presenting the results of
the electroweak corrections in terms of the fine structure
constant in the Thomson limit with $\alpha^{-1}=137.0359895$ and
the $Z$ mass $M_Z=91.1876$ GeV. The on-shell renormalisation
program, which we have described in detail
elsewhere\cite{nlgfatpaper}, uses $M_W$ as an input. However, the
numerical value of $M_W$ is derived through $\Delta
r$\cite{Hiokideltar} with $G_\mu=1.16639\times 10^{-5}{\rm
GeV}^{-2}$\footnote{The routine we use to calculate $\Delta r$ has
been slightly modified from the one used in our previous paper on
\eennht\cite{eennhletter} to take into account the new theoretical
improvements. It reproduces quite nicely the approximate formula
in \cite{deltarhollik-approx}. For the QCD coupling, we choose
$\alpha_S(M_Z)=0.118$.}. Thus, $M_W$ changes as a function of
$M_H$. For the  lepton masses we take $m_e=0.510999$ MeV,
$m_\mu=105.658389$ MeV and $m_\tau=1.7771$ GeV. For the quark
masses, beside the top mass $M_t=174$ GeV, we take the set
$M_u=M_d=63$ MeV, $M_s=94$ MeV\footnote{In \cite{eetthgrace}
$M_s=92$ MeV should read $M_s=94$ MeV.}, $M_c=1.5$ GeV and
$M_b=4.7$ GeV. With this we find, for example, that
$M_W=80.3766$GeV ($\Delta r=2.549\%$) for $M_H=120$ GeV and
$M_W=80.3477$GeV ($\Delta r=2.697\%$) for $M_H=180$ GeV.

As well known, from the direct experimental search of the Higgs
boson at LEP2, the lower bound of the \sm Higgs boson mass is
114.4 GeV\cite{mhlimit-direct}. On the other hand, indirect study
of the electroweak precision measurement suggests that the upper
bound of the \sm Higgs mass is about 200
GeV\cite{mhlimit-03-2003}. In this paper, we therefore only
consider a relatively light \sm Higgs boson and take the
illustrative values $M_H=120$GeV, $M_H=150$GeV and $M_H=180$GeV. In fact already
for $M_H>2M_W$ the process loses its interest for the LC as not
only the cross section decreases but also because the dominant
decay mode of the Higgs into $W$'s precludes a precision
study\cite{Baur3h-lhc-lc} of the triple Higgs coupling.

\subsection{Overview of the Feynman diagrams}
\begin{figure*}[htbp]
\begin{center}
\includegraphics[width=16cm,height=12cm]{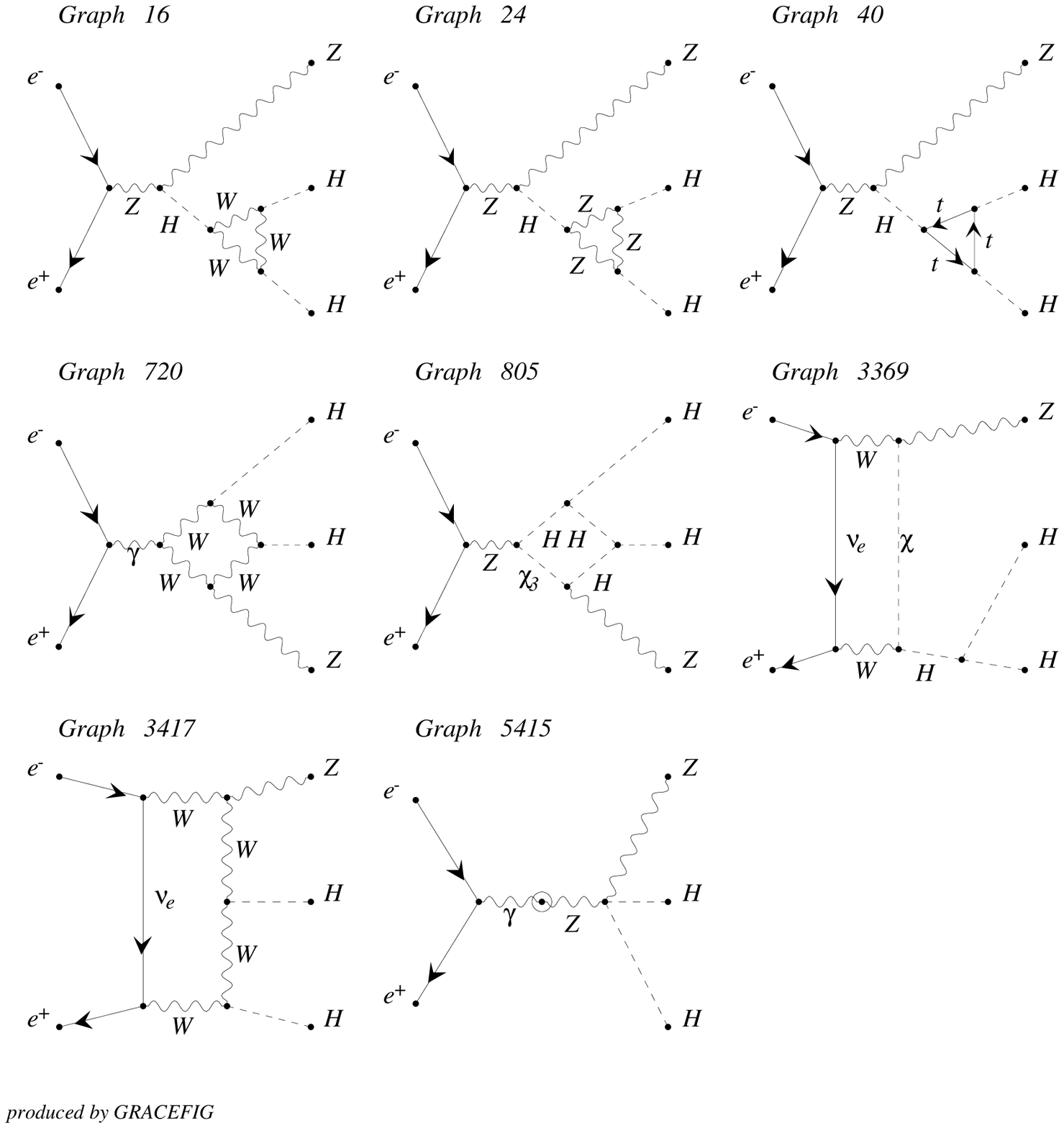}
\caption{\label{diagrams} {\small {\em A small selection of
different classes of loop diagrams contributing to \eezhht. We
keep the same graph numbering as that produced by the system. The
first three graphs represent corrections to the $HHH$ vertex. {\tt
Graph 720} and {\tt Graph 805} are box corrections representing
$Z^\star,\gamma^\star \ra HHZ$. {\tt Graph 3369} is a $t$-channel
induced diagram with $W^+ W^- \ra Z H^\star$ as a subprocess. {\tt
Graph 3417} is a typical pentagon diagram. {\tt Graph 5415} is a
two-point function correction. We do not show the QED graphs as
they amount to dressing the vertex $\epem Z^\star$.}}}
\end{center}
\end{figure*}

The full set of the Feynman diagrams within the non-linear gauge
fixing condition consists of 27 tree-level diagrams and as many as
5417 one-loop diagrams  for the electroweak ${\cal O}(\alpha)$
correction to \eezhh, see Fig.~\ref{diagrams} for a selection of
these diagrams. Neglecting the electron-Higgs coupling, the set of
diagrams still includes 6 tree-level diagrams and 1597 one-loop
diagrams. We define this latter set as the production set. To
obtain the results of the total cross sections, we use this
production set.

\section{Results}
\subsection{Tree-level}
\begin{figure*}[htbp]
\begin{center}
\includegraphics[width=16cm,height=9.5cm]{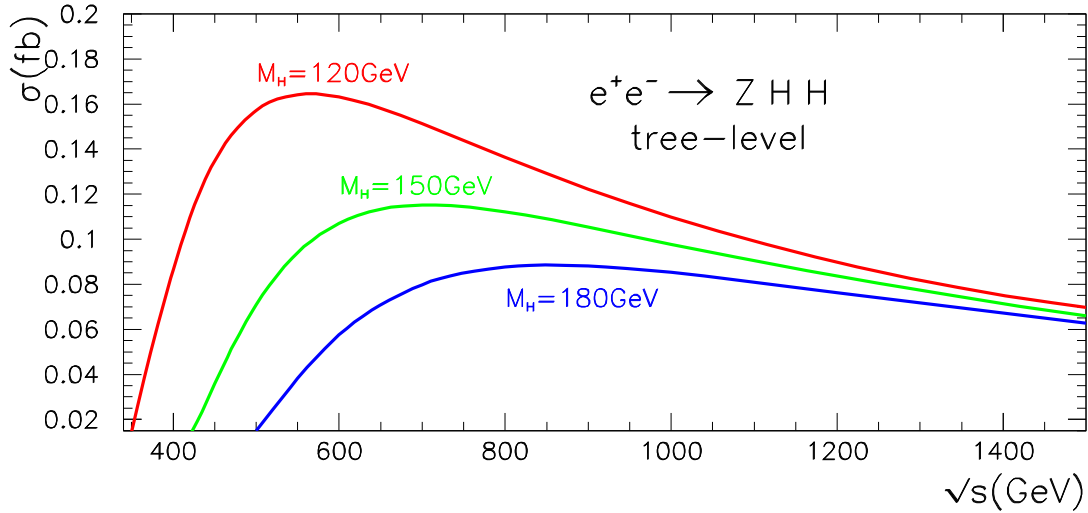}
\caption{{\em Total cross section for \eezhht as a function of the
centre of mass energy for $M_H=120,150,180$GeV.}}
\label{tree-level.fig}
\end{center}
\end{figure*}

To set the stage and to help understand the behaviour of the QED
corrections,  let us first give a brief summary on the cross
section at tree-level. This is shown in Fig.~\ref{tree-level.fig}.
Especially for the lightest Higgs mass, $M_H=120$GeV, there is a
sharp increase right after threshold. For this particular Higgs
mass the peak cross section is about $0.16$fb, at higher energies
the cross section decreases rather sharply. This peak cross
section decreases with increasing Higgs masses. Note, for further reference,
that for energies below $\sim 1.2$TeV there is a rather strong dependence on the
Higgs mass even in the restricted range of Higgs masses that we
are studying. For the measurement of the $HHH$ coupling it is most useful to run at the
maximum of the cross section. For a total integrated luminosity of
$1$ab$^{-1}$ the $1\sigma$ statistical error corresponds to about
a $2\%$ precision. Thus the theoretical knowledge of the cross
section at $0.2\%$ is more than sufficient. Expectedly at high
energies, the cross section shows little dependence on the Higgs
mass, at $\sqrt{s}=1.5$TeV, the cross section is about $0.06$fb.

\subsection{QED corrections: Slicing {\it vs} subtracting}
In our description of the computation of the one-loop virtual
electroweak corrections, the soft QED bremsstrahlung contribution
introduces the soft-photon cut-off parameter $k_c$. The $k_c$
dependence drops out when calculating the full \ordalft by
including the hard photon part, $d\sigma_H\equiv d\sigma_{\epem
\ra ZHH \gamma}(k_{\gamma}^0> k_c)$. With
$\delta_V=\delta_V^{EW}+\delta_V^{QED}$ the virtual loop
correction (including weak and QED corrections), the total
(integrated) \ordalft correction writes
\beqn
\sigma_{\ordalf}&=&\sigma_0(1+\delta_{\ordalf})=\int d\sigma_0
\left(1+\delta_V+\delta_S(k_c)\right)+\int d\sigma_H(k_c),\nonumber \\
&=&\underbrace{\int d\sigma_0 \left(1+\delta_V^{EW}\right)}_{\sigma_0(1+\delta_W)}+
\underbrace{\int d\sigma_0
\left(\delta_V^{QED}+\delta_S(k_c)\right)}_{\sigma_{V+S}^{QED}(k_c)}
+\underbrace{\int d\sigma_H(k_c)}_{\sigma_H(k_c)}\nonumber \\
&=& \sigma_0(1+\delta_W+\delta_{{\rm QED}}).
\eeqn
The second and third terms define our default slicing method.
The factorised soft-photon correction $d\sigma_S$ writes as
\beqn
d\sigma_S&=&d\sigma_0 \cdot \delta_S(k_c)=d\sigma_0 \cdot
f_{LL}(k_{\gamma}^0 < k_c)
\eeqn
with $d\sigma_0$ the tree-level $2\ra 3$ differential cross
section and the radiator function $f_{LL}$ is defined through
\beqn
f_{LL}&=&-e^2\;\int \frac{d^3k_{\gamma}}{(2 \pi)^3 2k_{\gamma}^0}
\left|\frac{p^-}{k_{\gamma}\cdot p^-}- \frac{p^+}{k_{\gamma}\cdot
p^+}\right|^2,
\eeqn
where $k_{\gamma}(p^{\pm})$ are photon ($e^{\pm}$) 4-momenta. As a
default in \grcl the hard photon contribution is performed by the
adaptive  Monte-Carlo integration {\tt BASES}\cite{bases} and the exact
matrix elements are generated by {\tt GRACE}. We then check that
within the Monte-Carlo integration errors there is no $k_c$
dependence, usually at the per-mil level or even better (see
below). For practically all $2\ra 3$ processes this is sufficient.

For the process at hand we have also introduced a second method
for the calculation of the hard part. A difficulty stems from the
large contribution from the collinear regions, ${\cal
O}(m_e^2/s)$, (of the hard photon radiation) which cancels a large
part of the negative contribution from
$\delta_{V+S}^{QED}=\delta_V^{QED}+\delta_S$. In order to have a
more stable result we use a subtraction
technique\cite{subtraction-grace} which is a variant of the dipole
subtraction introduced in \cite{dipoleqcd} for QCD and in
\cite{dipoleqed} for photon radiation. The idea is to add and
subtract to $d\sigma_H(k_c)$ a function that captures the leading
contribution of $d\sigma_H(k_c)$ and which is much easier to
integrate. In our case we use the function $d\tilde{\sigma}_0
\otimes f_{LL}$, where $d\tilde{\sigma}_0$ is derived from the
tree-level $d\sigma_0$ but with kinematics taking into account the
radiative photon emission. $\otimes$ stands for the phase-space
integration of the radiative photon convoluted with 3-body
phase-space of the tree-level cross-section. Therefore we
explicitly write
\beqn
\label{subtraction} \sigma_{\ordalf}&=& \int d\sigma_0
\left(1+\delta_V+\delta_S\right) + \int d\tilde{\sigma}_0 \otimes
f_{LL}(k_{\gamma}^0> k_c)\nonumber \\
&+& \int \left(d\sigma_H(k_{\gamma}^0>
k_c)-d\tilde{\sigma}_0\otimes f_{LL}(k_{\gamma}^0>
k_c)\right)\nonumber \\
&\equiv&\sigma_0 \left(1+\delta_W\right) +
\underbrace{\int \left(d\sigma_0 \;
\delta_V^{{\rm QED}}+d\tilde{\sigma}_0 \otimes f_{LL}\right)}_{\sigma^{QED}_A}
+\underbrace{\int \left(d\sigma_H(k_{\gamma}^0>
k_c)-d\tilde{\sigma}_0\otimes f_{LL}(k_{\gamma}^0>
k_c)\right)}_{\sigma^{QED}_B}\nonumber \\
\eeqn

\noi In $\sigma^{QED}_A$ of Eq.(\ref{subtraction}), the
convolution over the photon now includes both hard and soft photon
emission. The two-dimensional integration over the radiator
$f_{LL}$ is performed with the help of the {\em Good Lattice
Point} quasi Monte-Carlo method, see for example
\cite{latticepoint} for a description of this method. This ensures
appropriate cancellations of the singularities between the loop
and bremsstrahlung corrections at each point of (the tree-level)
phase-space. All integration over the remaining tree-level
differential cross section including both the virtual loop
corrections and photon emission is done with {\tt BASES}. For
$\sigma^{QED}_B$  we also use {\tt BASES} adapted to a $2\ra 4$
process. Here the subtraction means that all singularities are
made smooth at each point in phase space.

For the process at hand, which at tree-level proceeds through
$s$-channel $Z$-exchange, the virtual QED corrections form a gauge
invariant set which is all contained in the $e^+e^-Z^\star$
vertex. The dominant initial state QED virtual and soft
bremsstrahlung corrections are given by the universal soft photon
factor that leads to a relative correction\cite{eennhletter}
\beqn \label{dqeduniv}
\delta_{V+S}^{QED}=\frac{2 \alpha}{\pi}\left((L_e-1)\ln
\frac{k_c}{E_b}+\frac{3}{4}L_e + \frac{\pi^2}{6}-1 \right), \quad
\; L_e=\ln(s/m_e^2) \;,
\eeqn
\noi where $m_e$ is the electron mass, $E_b$ the beam energy
($s=4E_b^2$) and $k_c$ is the cut on the soft photon energy.

Although this approach of extracting the full QED correction is
the most simple one, we have also calculated the full QED
corrections separately  and subtracted their contributions from
the full ${\cal O}(\alpha)$. In order to perform this subtraction,
the QED virtual corrections are generated by dressing the
tree-level diagrams with one-loop photons (the photon self-energy
is not included in this class). Moreover one needs to include some
counterterms. One only has to take into account the purely
photonic contribution to  the wave function renormalisation
constants of the electron. Performing this more direct computation
is another test on the system.
\begin{table}[hbt]
\begin{center}
\noindent
\begin{tabular}{|r||c||r|c|r||c|c|r|}
\hline \multicolumn{1}{|c||}{\(\sqrt{s}\)} &
\multicolumn{1}{c||}{$\sigma_{\rm tree}=\int d\sigma_0$} & $\sigma^{QED}_A$ & \multicolumn{1}{c|}{$\sigma^{QED}_B$} &
\(\delta_{QED}^\prime\) &
$\int d\sigma_{V+S}^{QED}(k_c)$    & \multicolumn{1}{c|}{$ \int d\sigma_H(k_c) $}   & \(\delta_{QED}\) \\
\multicolumn{1}{|c||}{[GeV]} & \multicolumn{1}{c||}{[fb]} &
\multicolumn{1}{c|}{[fb]} & \multicolumn{1}{c|}{[fb]}
    & \multicolumn{1}{c||}{[\%]} &
\multicolumn{1}{c|}{[fb]} & \multicolumn{1}{c|}{[fb]}
    & \multicolumn{1}{c|}{[\%]} \\
\hline $ 400$ & $0.08644$ & $-0.01636$& $0.00007$ & $-18.84$
    & $-0.11971$ & $0.10344$ & $-18.83$ \\
$ 600$ & $0.16312$ & $-0.00743$ & $0.00155$ & $ -3.61$
    & $-0.24122$ & $0.23535$ & $ -3.60$ \\
$ 800$ & $0.13631$ & $-0.00075$& $0.00284$ & $  1.53$
    & $-0.21091$ & $0.21300$ & $  1.53$ \\
$1000$ & $0.10985$ & $ 0.00130$& $0.00344$ & $  4.32$
    & $-0.17594$ & $0.18067$ & $  4.31$ \\
$1500$ & $0.06983$ & $ 0.00201$& $0.00360$ & $  8.02$
    & $-0.11887$ & $0.12448$ & $  8.04$ \\
\hline
\end{tabular}
\caption{\label{tab-slice-subtract}{\em  Comparison between the
results of the subtraction method and the slicing method as
described in the text for totally integrated quantities.
$M_H=120$GeV and $k_c=10^{-3}$GeV.
$\delta_{QED}^\prime$($\delta_{QED}$)
 is the percentage total QED correction in the subtraction (slicing) method.}}
\end{center}
\end{table}

Table~\ref{tab-slice-subtract} shows a comparison of the total QED
correction in our default slicing method and our numerical
implementation of the subtraction method. We see that the
agreement is excellent (at the level of $10^{-4}$ accuracy). The
table also makes clear the advantage of the subtraction method in
that no large cancellation between the part containing the virtual
contribution and the rest occurs, moreover both are much smaller
than the two parts obtained in the slicing method. There,  as
advertised, a large cancellation takes part. We note also  that
$\int d\sigma_{V+S}^{QED}(k_c)$ in Table~\ref{tab-slice-subtract},
is in excellent agreement with the result obtained by multiplying
the tree-level cross section with the universal factor of
Eq.(\ref{dqeduniv}). It is important to note that the full QED
correction are rather large and negative around threshold,
moderate around the peak and increase steadily for high energies.
Most of this can be explained from the behaviour of the tree-level
cross section, see Fig.~\ref{tree-level.fig}, as a boost  (sort of
a radiative return) towards lower energies.

\subsection{The genuine weak corrections}
\begin{figure*}[bhtp]
\begin{center}
\leavevmode
\includegraphics[width=16cm,height=7cm]{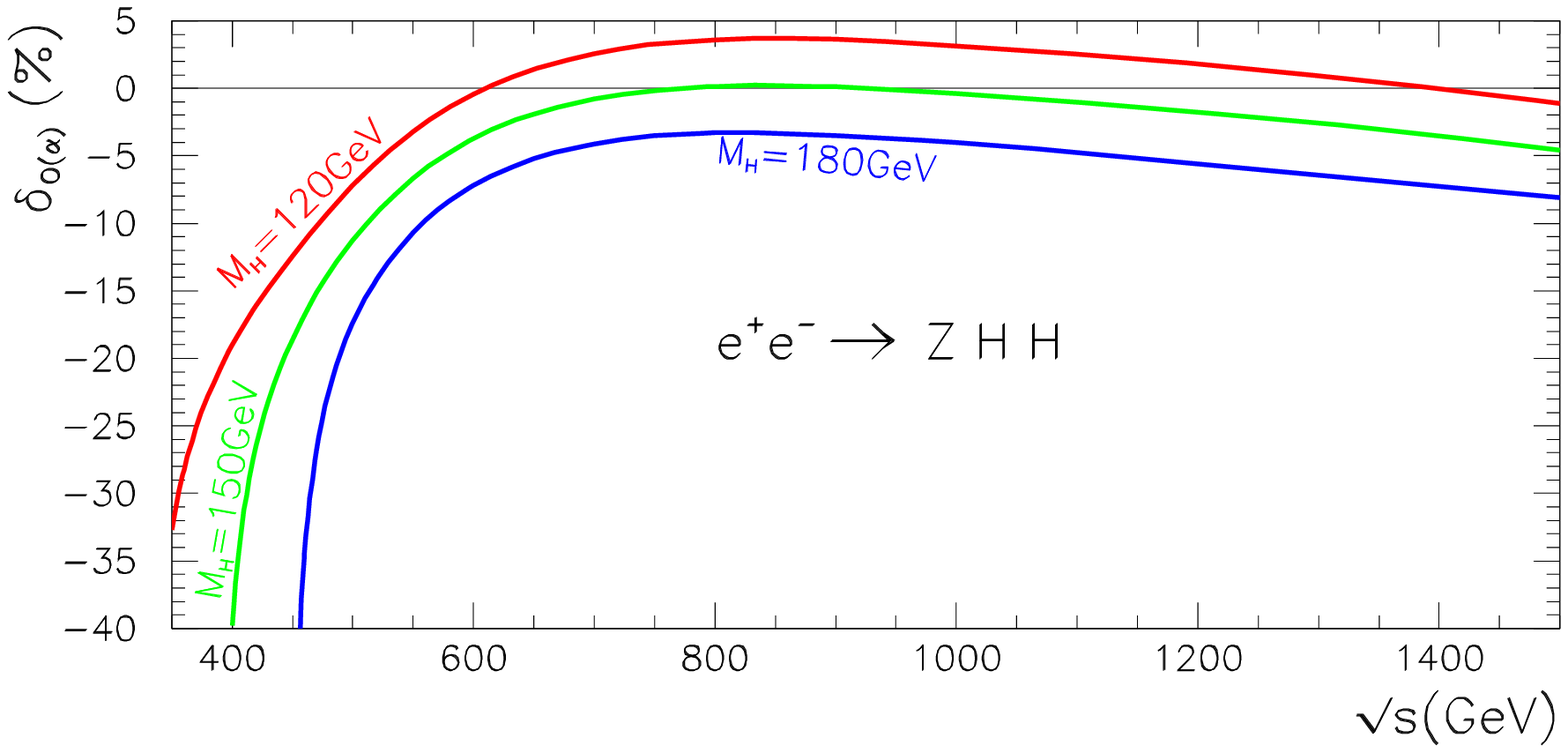}
\mbox{\hspace*{-0.3cm}\mbox{\includegraphics[width=16cm,height=10cm]{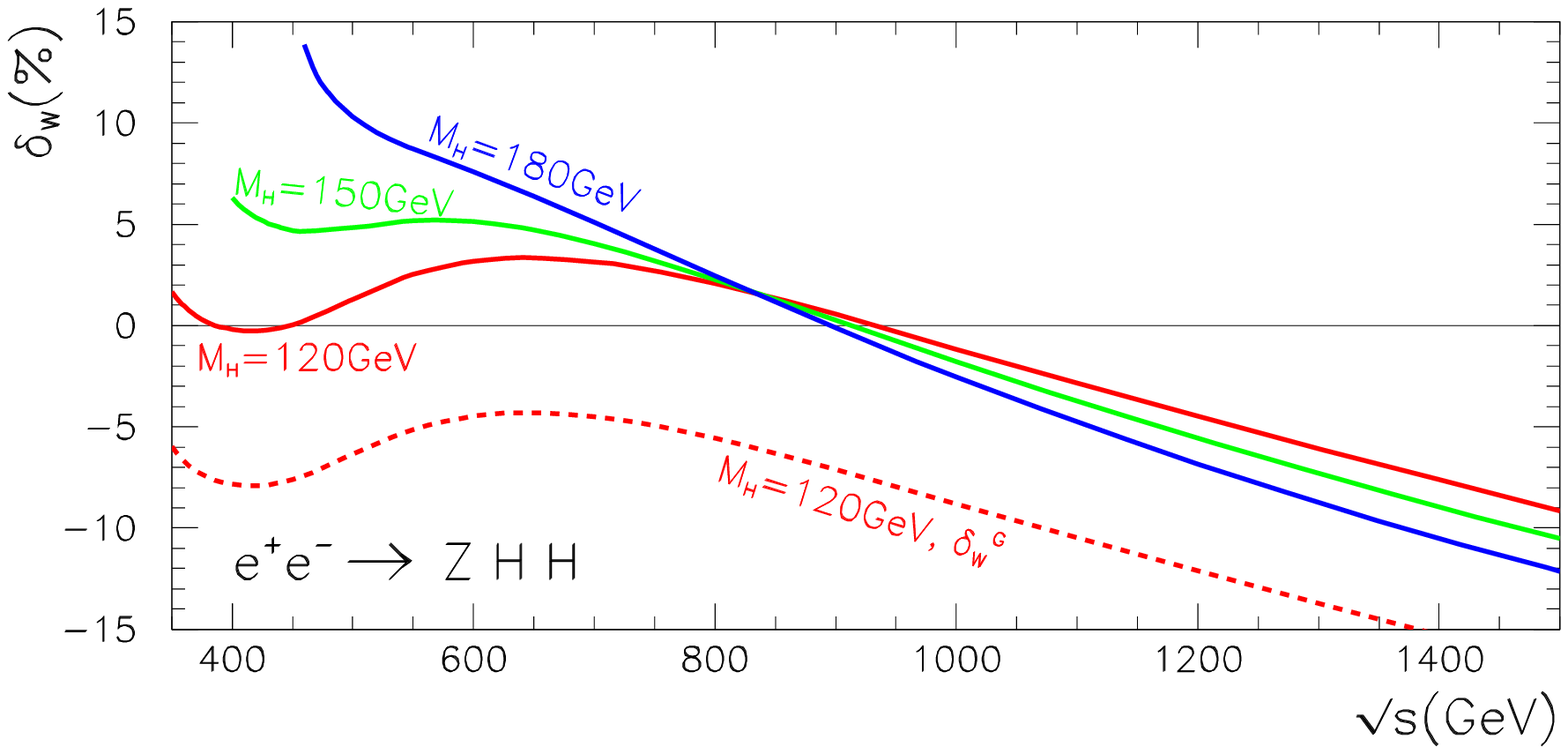}}
\raisebox{5.5cm}{ \mbox{ \hspace*{-8.5cm}
\includegraphics[width=7.5cm,height=4.cm]{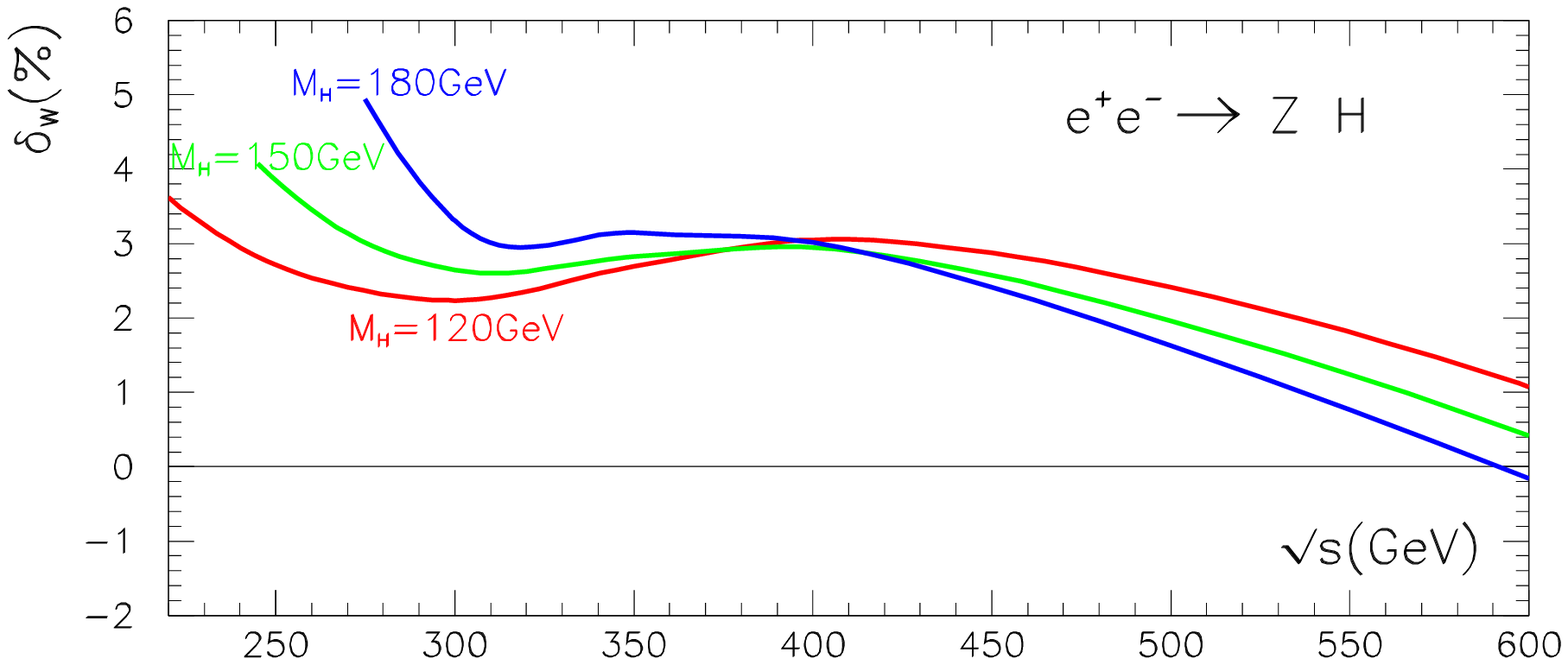}}}}
\caption{\label{delrc.fig}{\em The full ${\cal O}(\alpha)$
relative correction (top panel) and the relative electroweak
correction $\delta_W$ (lower panel) as a function of energy for
$M_H=120,150,180$GeV. In addition, the genuine weak correction
$\delta_W^{G}$ in the $G_\mu$ scheme is presented for
$M_H=120$GeV (dotted line). The insert shows $\delta_W$ for $\epem \ra ZH$ for the same
three Higgs masses.}}
\end{center}
\end{figure*}
We now turn to the genuine weak corrections which are the most
interesting from a physics point of view. We have already
discussed how in this process these corrections can be
unambiguously defined. The total \ordalft correction and the genuine
weak corrections as a function of energy for our three representative
Higgs masses are shown in Fig.~\ref{delrc.fig}.

\begin{table*}[hbt]
\begin{center}
\begin{tabular}{|c|c|c|c|r|r|}
\hline \rule[-1mm]{0mm}{5mm} $\sqrt{s}$ & $M_H$(GeV)&
$\sigma_{tree}$(fb) & $\Delta\sigma_{{\cal O}(\alpha)}$(fb)&
$\delta_{{\cal O}(\alpha)}(\%)$ &
$\delta_W(\%)$\\
\hline \rule[-1mm]{0mm}{5mm} 600 GeV & 120 &
$1.6311(2)\times10^{-1}$ & $-0.719(3)\times10^{-3}$ &
$-0.44(1)$ & $3.17(1)$ \\
\rule[-1mm]{0mm}{5mm} ~       & 150 & $1.0704(1)\times10^{-1}$ &
$-3.987(3)\times10^{-3}$ &
$-3.72(1)$ &  $5.14(1)$ \\
\rule[-1mm]{0mm}{5mm} ~       & 180 & $5.7491(5)\times10^{-2}$ &
$-4.136(3)\times10^{-3}$ &
$-7.19(1)$ &  $7.60(1)$ \\
\hline \rule[-1mm]{0mm}{5mm} 800 GeV & 120 &
$1.3631(1)\times10^{-1}$ & $+4.911(6)\times10^{-3}$ &
$ 3.60(1)$ &  $2.07(1)$ \\
\rule[-1mm]{0mm}{5mm} ~       & 150 & $1.1221(1)\times10^{-1}$ &
$-0.167(8)\times10^{-3}$ &
$0.14(1)$ & $2.27(1)$ \\
\rule[-1mm]{0mm}{5mm} ~       & 180 & $8.7604(8)\times10^{-2}$ &
$-2.917(6)\times10^{-3}$ &
$-3.33(1)$ & $2.45(1)$ \\
\hline \rule[-1mm]{0mm}{5mm} 1 TeV  & 120 &
$1.0986(1)\times10^{-1}$ & $+3.462(6)\times10^{-3}$ &
$ 3.15(1)$ &  $-1.17(1)$ \\
\rule[-1mm]{0mm}{5mm} ~       & 150 & $9.7667(9)\times10^{-2}$ &
$-0.386(8)\times10^{-4}$ &
$-0.40(1)$ & $-1.78(1)$\\
\rule[-1mm]{0mm}{5mm} ~      & 180 & $8.5245(8)\times10^{-2}$ &
$-3.413(9)\times10^{-3}$ &
$-4.00(1)$ & $-2.54(1)$ \\
\hline
\end{tabular}
\caption{{\em  Total ${\cal O}(\alpha)$ and genuine weak
corrections. }} \label{ordalptab}
\end{center}
\end{table*}

Specialising first to $M_H=120$GeV, the genuine weak corrections
expressed in the $\alpha$-scheme are very small at threshold and
also for energies corresponding to the peak cross section. At
threshold this is in sharp contrast to the full \ordalft cross
section which is overwhelmingly dominated by the pure QED
correction. For the range of the most interesting centre of mass
energies where the cross section is largest, the (relative) weak
correction is never more than $4\%$. Notice also the ``spoon-like"
behaviour of the weak correction at the lowest energies. This also
occurs in $\epem \ra ZH$ as displayed in the insert of
Figure~\ref{delrc.fig}\footnote{This calculation of the
electroweak correction to $\epem \ra ZH$ has also been performed
by {\tt GRACE}.} and has also been observed in the so-called
$s$-channel of $\epem \ra \nu \bar \nu H$ (see Fig.~3a
of\cite{eennhletter}). This seems to be due to a competition
between the bosonic and fermionic contributions at energies close
to threshold but a more detailed investigation is needed to
confirm the origin of this common feature. Passed the peak, the
weak correction steadily turns negative reaching about $-10\%$ at
$1.5$TeV. Comparing to the case with $M_H=150$GeV and
$M_H=180$GeV, we see that, in fact, past $\sqrt{s}=800$GeV, the
Higgs mass dependence of the genuine electroweak correction is
small. Larger differences between the 3 Higgs masses considered
here only appear for the smallest energies around threshold. Like
for $M_H=120$GeV, for both $M_H=150$ and $M_H=180$GeV the weak
correction around the peak is also within about $5\%$ and thus
well contained. Note that the decrease of the weak correction as
the energy increases is faster with the higher Higgs mass, a trend
which is similar to that inherited from $\epem \ra ZH$ (see
insert of Figure~\ref{delrc.fig}). \\
\noi Having subtracted the genuine weak corrections one could also
express the corrections in the $G_\mu$ scheme by further
extracting the rather large universal weak corrections that affect
two-point functions through $\Delta r$. This defines the genuine
weak corrections in the $G_\mu$ scheme as
$\delta_W^{G}=\delta_W-3\Delta r$. For \eennht this procedure
helps absorb a large part of the weak corrections. Another
advantage is that much of the (large) dependence due to the light
fermions masses also drops out. However one sees,
Fig.~\ref{delrc.fig}, that  especially for high energies, this
scheme fails to properly encapsulates the bulk of the radiative
corrections.  This is akin to what happens in
\eezht\cite{eezhsmrc} and $\epem \ra t \bar t H$\cite{eetthgrace}
where the bosonic weak corrections become important (and negative)
at high energies. \\
\noi To summarise it is perhaps worth to stress
that if the total cross section even at its peak value would not
be measured better than $10\%$ it would then be difficult to
observe the genuine weak corrections. One should also observe that
the approximate leading top mass correction to the Higgs
self-couplings as given in Eq.~(\ref{mt4hhh}) and which has no energy
dependence can not reproduce the bulk of the weak correction.

\subsection{Corrections to the $M_{HH}$ distribution}
\begin{figure*}[hbtp]
\begin{center}
\includegraphics[width=10cm,height=9.5cm]{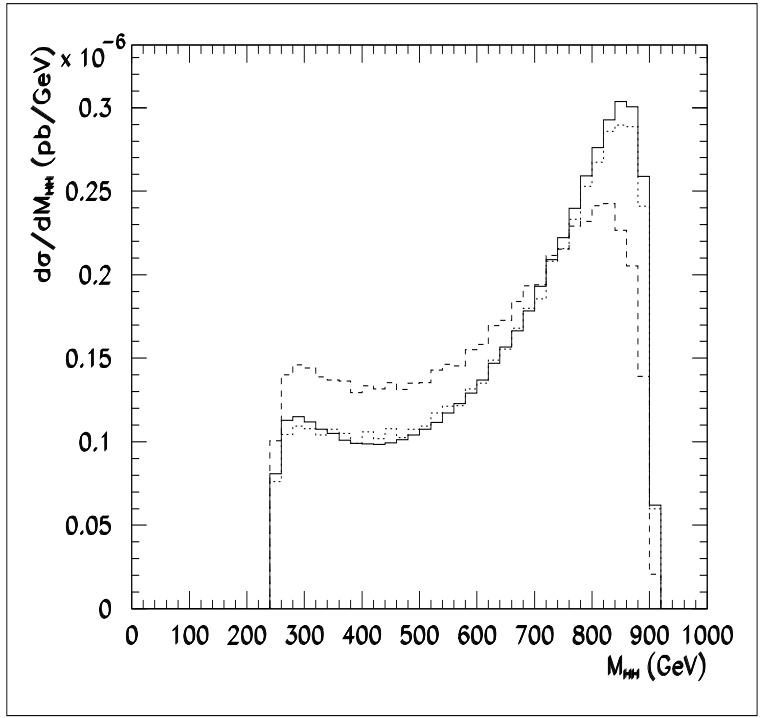}
\caption{\label{mhdiss.fig}{\em $d\sigma/d M_{HH}$ for
$M_H=120$GeV at $\sqrt{s}=1$TeV. We show the tree-level (full
curve), the effect of including only the genuine weak corrections
(dotted curve) and the effect of including  the full \ordalft
(dashed curve). }}
\end{center}
\end{figure*}

The $M_{HH}$ distribution has been
shown\cite{hhhbattagliaboos,jlc-3h} to be a good discriminator for
isolating the $HHH$ vertex, taking advantage of the fact that the
two final Higgs mimic the decay product of a scalar (the virtual
Higgs in the $HHH$ vertex)\cite{ChopinHiggs}. It is therefore
important to enquire how this distribution gets affected by loop
corrections. We have chosen as an illustrative case, $M_H=120$GeV
at a centre of mass energy $\sqrt{s}=1$TeV which is the same
configuration that was studied recently in the simulation of
\cite{jlc-3h} which included, at tree-level, the effect of an
anomalous triple Higgs coupling. In there it was shown that a
deviation in this coupling affects primarily the lower end of the
$M_{HH}$ spectrum, whereas the higher end is little affected. This
is rather similar to the effect of the full \ordalft correction as
shown in Fig.~\ref{mhdiss.fig} especially at the lower edge of the
spectrum. The full \ordalft correction, which for the integrated
cross section reaches about $+3\%$, shows in fact a depletion from
the higher $M_{HH}$ to the lower  $M_{HH}$ values, again an effect
due to the radiative return. After extracting the QED correction,
the distribution including the weak corrections is hardly,
considering the foreseen precision, noticeable though. Therefore
we conclude that an anomalous coupling could still be
distinguished, if large enough, in this distribution provided a
proper inclusion of the initial QED corrections is allowed for in
the experimental simulation.

\section{Conclusions}

We have performed a full one-loop correction to the process \eezhh.
This is a process which is interesting and of importance mainly
because it gives access to the measurement of the triple Higgs coupling.
Our calculation shows that especially not far from threshold the QED
corrections are large so that a proper resummation of the initial
state radiation needs to be performed. However in this energy range
the cross section are modest and probably not measurable. At
energies where the cross section is largest on the other hand, the
corrections are modest especially for the lightest Higgs mass of
$M_H=120$GeV. This applies also to the genuine weak corrections at
around the peak of the cross sections. Indeed in these regions
these genuine weak corrections do not exceed $\sim 5\%$ and are therefore
below the expected experimental precision. We have also investigated how
these corrections were distributed as a function of the
discriminating variable $M_{HH}$, the invariant mass of the Higgs pair,
having in mind the use of  this variable for the  extraction of the triple
Higgs vertex. We find
that the genuine weak corrections, contrary to the QED corrections,
hardly affect the shape of the distribution at least for energies
where this distribution is to be exploited.

\vspace{1cm}
\noi {\bf Acknowledgments} \\
\noi This work is part of a collaboration between the {\tt GRACE}
project in the Minami-Tateya group and LAPTH. We would like to thank
D. Perret-Gallix for his continuous interest and encouragement.
This work was supported in part by the Japan Society for Promotion
of Science under the Grant-in-Aid for scientific Research
B(N$^{{\rm o}}$ 14340081) and PICS 397 of the French National
Centre for Scientific Research (CNRS).

\newpage
{\bf \large Note added}

While  finalising this paper a calculation of the same process
appeared\cite{eezhhchinese}. We have run our program with the same
set of input parameters and compared the total \ordalft results,
see Table.~\ref{chinese-compare}. We find a very good
agreement, within $0.1\%$, for centre of mass energies up to $\sqrt{s}=800$GeV for
all three Higgs masses, $M_H=115,150,200$GeV.  As the
energy increases the agreement worsens  somehow, at
$\sqrt{s}=1.5$TeV the agreement is within $0.3\%$ but at $\sqrt{s}=2$TeV the agreement
is no better than $0.8\%$.
\small

\begin{table}[hbt]
\newcommand{\phm}{\phantom{-}}
$$\begin{array}{c@{\quad}c@{\quad}l@{\quad}l@{\quad}l}
\hline \sqrt{s}~ ({\rm GeV}) & M_H~ ({\rm GeV}) & \sigma_{{\rm
tree}}~ ({\rm fb}) & \sigma_{\ordalf}~ ({\rm fb}) &
\delta_{\ordalf}~ [\%] \\
\hline
    & 115 & {\it 0.17493(2)} &{\it 0.1629(2)} & {\it -6.9(1)} \\
    &  &0.17491(2) & 0.16282(2) & -6.91(1)\\
500 & 150 &{\it 0.071834(6)} &{\it 0.06357(6)} & {\it -11.50(7)} \\
    &  & 0.071830(5) & 0.063529(9)& -11.59(9)\\
    & 200 &{\it 0.49611(3) \cdot 10^{-3} }& {\it0.3329(2) \cdot 10^{-3}} &{\it -32.90(4)} \\
    &  & 0.49606(4) \cdot 10^{-3}& 0.332(3) \cdot 10^{-3}& -33.0(6)\\
\hline
    & 115 & {\it 0.14156(3)} &{\it 0.1471(3)} &{\it +3.9(2) }\\
    &  & 0.14155(1) & 0.14705(2) & +3.89(1)\\
800 & 150 & {\it0.11363(2) }&{\it 0.1135(2) }&{\it -0.1(2)} \\
&  & 0.11362(1) & 0.11353(1) & -0.08(7)\\
    & 200 &{\it 0.07246(1) }&{\it 0.0705(1)} &{\it -2.7(1)} \\
    &  & 0.072454(7)& 0.07044(1) & -2.78(1)\\
\hline
     & 115 &{\it 0.07119(2)} &{\it 0.0704(3)} &{\it -1.1(4) }\\
     &  & 0.07118(1)& 0.07058(2) & -0.85(3)\\
1500 & 150 & {\it 0.06684(2) }&{\it 0.0634(2)} &{\it -5.1(3)} \\
&  & 0.06683(1) & 0.06359(2) & -4.86(3)\\
& 200 & {\it 0.06165(1) }&{\it 0.0569(2) }&{\it -7.7(3)} \\
&  & 0.061644(6) & 0.05707(2) & -7.42(3)\\
\hline
     & 115 &{\it 0.05021(1)} & {\it0.0473(2)} &{\it -5.8(4) }\\
     &  & 0.05021(1)& 0.04773(2) & -4.95(4)\\
2000 & 150 &{\it 0.04812(1) }&{\it 0.0435(2)} &{\it -9.6(4)} \\
&  & 0.048119(5) & 0.04387(3)&  -8.83(7)\\
     & 200 &{\it 0.04630(1) }&{\it 0.0408(2) }& {\it-11.9(4)} \\
     &  & 0.046300(4) & 0.04115(3) & -11.13(6)\\
\hline
\end{array}$$
\caption{{\em Comparison of results between \cite{eezhhchinese} and our
code. The results of \cite{eezhhchinese} are in italic.
The Born cross section $\sigma_{{\rm tree}}$, the full order ${\cal
O}(\alpha)$  corrected
cross section $\sigma_{\ordalf}$ and the full  \ordalft relative correction $\delta_{\ordalf}$ for various Higgs boson masses and centre of mass
energies are compared.}}
\label{chinese-compare}
\end{table}


 \vspace{1cm}

\normalsize

\newpage

\end{document}